# Multi-Stage Holomorphic Embedding Method for Calculating the Power-Voltage Curve

Bin Wang, *Student Member, IEEE*, Chengxi Liu, *Member, IEEE*, Kai Sun\*, *Senior Member, IEEE*

*Abstract*—The recently proposed non-iterative load flow method, called the holomorphic embedding method, may encounter the precision issue, i.e. nontrivial round-off errors caused by the limit of digits used in computation when calculating the power-voltage (P-V) curve for a heavily loaded power system. This letter proposes a multi-stage scheme to solve such a precision issue and calculate an accurate P-V curve. The scheme is verified on the New England 39-bus power system and benchmarked with the result from the traditional continuation power flow method.

*Index Terms*—Holomorphic embedding method (HEM), multi-stage HEM, holomorphic error embedding method, continuation power flow, power-voltage curve.

## I. INTRODUCTION

THE power voltage (P-V) curve is a useful tool for steady state voltage stability analysis, whose calculation usually involves solving a series of power flow problems. The traditional continuation power flow method (CPF) [1] adopts the Newton-Raphson (N-R) method in its predictor-corrector scheme, where, although the divergence is usually not encountered, each step still has to solve the power flow equations by a number of iterations. In the last decade, some non-iterative methods for calculating power flow equations have been proposed and developed [2], among which the Holomorphic Embedding Method (HEM) firstly proposed in [3] has the capability of handling large-scale power systems [4]. However, the HEM has the precision issue that any limited digits used in the computations introduce nontrivial round-off errors, which finally reduce the accuracy of power flow solutions. The root cause of this issue can be explained by the Stahl's theory that the Padé matrix used in an HEM will have a larger condition number when the system approaches the nose point, which then slows down the convergence rate of the HEM method such that the computations using limited digits will inevitably introduce non-ignorable errors. Although adding more digits in the computations or adopting the Padé approximants may improve the accuracy, still, the precision issue cannot be eliminated with limited digits [5].

This letter proposes a multi-stage HEM (MSHEM) adopting a prediction-correction scheme to calculate the P-V curve with any pre-defined error tolerance. A new holomorphic error embedding method is proposed as the corrector. Thus, unlike the CPF and a traditional HEM, the MSHEM not only keeps the non-iterative advantage but also has higher precision.

## II. FOUR TYPICAL HEMs

Four typical HEMs are summarized in Table I and numbered as M1-M4, where the second to fourth columns respectively indicate whether each method can deal with systems having PV buses, whether each method uses a physical germ, and what specific way of embedding is used.

Note that there are several other HEMs implemented in the literatures, which may differ in the modeling of PV buses, the embedding on load buses and the ways to refine the resultant Taylor expansions in complex variable *s*. Since the multi-stage scheme can be similarly developed for them, they will not be mentioned or referred to due to the limited space.

TABLE I
FOUR TYPICAL HEMs

| Method | Consider PV buses? | Use a physical germ\*? | Embedding on PQ buses\*\* | Reference |
|--------|--------------------|------------------------|---------------------------|-----------|
| M1 | No | No | $s \cdot (P\text{-}jQ)$ | [3] |
| M2 | Yes | No | $s \cdot (P\text{-}jQ)$ | [5] |
| M3 | Yes | Yes | $s \cdot (P\text{-}jQ)$ | [6] |
| M4 | Yes | Yes | $(1+k^*s) \cdot (P\text{-}jQ)$ | [7] |

\*The physical germ is a germ solution that the entire resulting voltage solution of any PQ bus, i.e. $V(s)$ for any $s$, always lies on the actual P-V curve of the system. As a comparison, the germ solution used in M1 and M2 is not a physical germ, since only one point, i.e. $V(s)$ at $s=1$, lies on the actual P-V curve.
\*\*The embedding in M1-M3 only allows the load to increase from zero to the target loading condition, while the embedding in M4 allows the load to increase\decrease from a certain loading condition.

## III. PROPOSED MULTI-STAGE SCHEME FOR HEMs

As a comparison, the traditional CPF calculates the P-V curve starting from a certain power flow solution, linearly predicting the adjacent points on the P-V curve step by step and correcting these points by the N-R method. The proposed MSHEM also starts from a certain power flow solution but predicts points on the P-V curve nonlinearly, so it has larger steps than the CPF as illustrated in Fig.1.

This work was supported by the NSF CURENT engineering research center.
B. Wang, C. Liu and K. Sun\* are with the University of Tennessee, Knoxville, TN, 37996 USA. (e-mail: {bwang, cliu48, kaisun}@utk.edu).

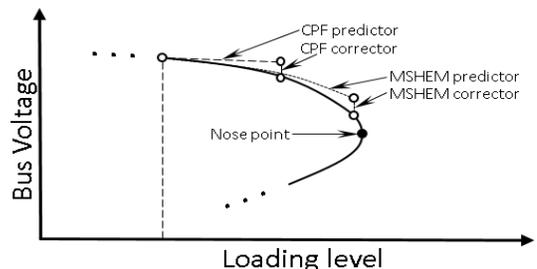

Fig.1. One prediction-correction step of the CPF and of the proposed MSHEM

**Predictor**: the MSHEM uses the voltage solution functions from HEMs [6][7], i.e. a nonlinear function $V(s)$ of $s$, to predict the next point on the P-V curve.

**Step length**: For each stage of the MSHEM, the step length is determined by finding a point $s$ such that the error of $V(s)$ reaches a pre-specified tolerance of power-flow equations.

**Corrector**: Although the N-R method can also be adopted here as a corrector, a new non-iterative method based on holomorphic embedding called **holomorphic error embedding (HEE)** method is proposed below to make calculation of the entire P-V curve free of iterations. Different from the existing HEMs, which connect solved power flows, i.e. the germ solution, to the target solution, the proposed HEE method connects an inaccurate power flow solution to an accurate one.

Let the last point of a certain stage be $\mathbf{x}_l$. Since $\mathbf{x}_l$ is an inaccurate solution from the prediction, power-flow equations $\mathbf{f}(\mathbf{x}_l)=\mathbf{0}$ are not completely satisfied. Let the error vector be $\boldsymbol{\varepsilon}$.

$$\boldsymbol{\varepsilon} = \mathbf{f}(\mathbf{x}_l) \neq \mathbf{0} \tag{1}$$

Define $\mathbf{g}$ in the complex domain as a vector function of $s$.

$$\mathbf{g}(s) = \mathbf{f}(\mathbf{x}(s)) - (1-s) \cdot \boldsymbol{\varepsilon} \tag{2}$$

$$\mathbf{x}(s) = \mathbf{x}_l + \sum_{k=1}^{\infty} \mathbf{x}_{lk} \cdot s^k \tag{3}$$

where $\mathbf{x}_{lk}$ are unknown parameter vectors. Then, we have the following three observations:

(i) $s=0$ gives a germ solution $\mathbf{x}_l$, since $\mathbf{g}(0) = \mathbf{f}(\mathbf{x}_l) - \boldsymbol{\varepsilon} = \mathbf{0}$ by definition.

(ii) $s=1$ leads to the target solution, since $\mathbf{g}(1) = \mathbf{f}(\mathbf{x}(1)) = \mathbf{0}$. Then, $\mathbf{x}(1)$ is the accurate power flow solution.

(iii) Similar to the traditional HEMs, those unknown parameter vectors $\mathbf{x}_{lk}$ can be determined recursively by equating the coefficients of $s$ on both sides of (2) and solving the resulting linear equations.

The correction is necessary when the error in power flow solutions introduced by using only limited digits becomes non-ignorable, e.g. when the system approaches the nose point.

**Convergence**: the MSHEM stops when the next step length is smaller than a pre-specified value, e.g. 1MW.

**Precision issue**: the MSHEM does not have the precision issue because even a limited number of digits are used, e.g. 16 digits, the step length can always be automatically adjusted to give the accurate P-V curve satisfying a pre-specified tolerance.

**Selection of HEMs**: Only HEMs using physical germs, e.g. M3 and M4, can be used for the MSHEM since all points of the solution $V(s)$ need to be actual points on the P-V curve of the original system, rather than the only point $V(s=1)$ in M1 and M2.

## IV. Test on New England 39-Bus Power System

The effectiveness of the proposed MSHEM is tested on the New England 39-bus power system. All loads and active power output of generators are proportionally increased from the current loading condition. P-V curves calculated by the CPF are benchmarks, where the tolerance for power flow calculation is set to $10^{-8}$. The HEM and HEM using 16 digits are tested, where the tolerance of the HEE method is set to $10^{-8}$. Fig.2a contains the P-V curves from all buses while the P-V curve of bus 8, which is in a load center area, is presented in Fig.2b. The largest power mismatch of power-flow equations using solution points on the P-V curves by different methods are shown in Fig.3. It is worth mentioning that the CPF uses 228 steps to approach the nose point, while the proposed MSHEM takes only 3 steps.

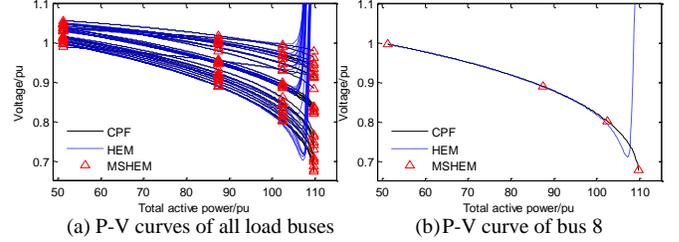

(a) P-V curves of all load buses    (b) P-V curve of bus 8

Fig.2. P-V curves respectively calculated by CPF, HEM and MSHEM.

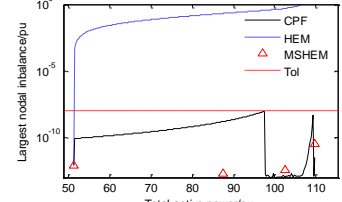

Fig.3. Largest mismatches of power balance equations.

## V. Conclusion

This letter proposes a multi-stage holomorphic embedding method (MSHEM) to enable accurate HEM-based P-V curve calculation. The proposed MSHEM is free of iterations, which uses multiple stages to avoid the precision issue encountered by existing HEMs when dealing with heavily loaded power systems. In each stage, the voltage solution functions $V(s)$ from HEMs with physical germs are used as the predictor and the holomorphic error embedding (HEE) method is proposed to be the corrector. Test results on the New England 39-bus power system show that the proposed MSHEM is able to calculate P-V curves satisfying the pre-specified tolerance within very few steps.

In addition, the proposed MSHEM is able to provide piecewise analytical expressions of a P-V curve due to the merits of HEMs. Such information could be useful for predicting the limit violations of line flows and generator outputs.